\documentclass[amsmath,amssymb,a4paper,preprint,prl,showpacs]{revtex4}
\usepackage{amssymb}
\usepackage{txfonts}
\usepackage{graphicx}
\usepackage{ulem}
\usepackage{color}

\begin{document}

\title{Model for Topological Phononics and Phonon Diode}
\author{Yizhou \surname{Liu}$^{1,2,3}$}
\author{Yong \surname{Xu}$^{1,2,3}$}
\email{yongxu@mail.tsinghua.edu.cn}
\author{Shou-Cheng \surname{Zhang}$^{4,5}$}
\author{Wenhui \surname{Duan}$^{1,2,5}$}
\email{dwh@phys.tsinghua.edu.cn}

\affiliation{
$^1$State Key Laboratory of Low Dimensional Quantum Physics, Department of Physics, Tsinghua University, Beijing 100084, People's Republic of China \\
$^2$Collaborative Innovation Center of Quantum Matter, Beijing 100084, People's Republic of China \\
$^3$RIKEN Center for Emergent Matter Science (CEMS), Wako, Saitama 351-0198, Japan \\
$^4$Department of Physics, McCullough Building, Stanford University, Stanford, California 94305-4045, USA \\
$^5$Institute for Advanced Study, Tsinghua University, Beijing 100084, People's Republic of China}

\begin{abstract}
The quantum anomalous Hall effect, an exotic topological state first theoretically predicted by Haldane and recently experimentally observed, has attracted enormous interest for low-power-consumption electronics. In this work, we derived  a Schr{\"o}dinger-like equation of phonons, where topology-related quantities, time reversal symmetry and its breaking can be naturally introduced similar as for electrons.
Furthermore, we proposed a phononic analog of the Haldane model, which gives the novel quantum (anomalous) Hall-like phonon states characterized by one-way gapless edge modes immune to scattering. The topologically nontrivial phonon states are useful not only for conducting phonons without dissipation but also for designing highly efficient phononic devices, like an ideal phonon diode, which could find important applications in future phononics.
\end{abstract}


\maketitle

The discovery of the quantum spin Hall (QSH) effect and topological insulators (TIs)~\cite{kane2005,bernevig2006,konig2007} that are new states of quantum matter has revolutionized the research of condensed matter physics and material science~\cite{qi2010,hasan2010,qi2011}. The quantum anomalous Hall (QAH) effect, a quantum Hall (QH) effect without Landau levels, represents another type of topological state with broken time reversal symmetry (TRS). It was theoretically predicted by Haldane in 1988~\cite{haldane1988} and has been experimentally observed in a magnetic-doped TI recently~\cite{chang2013}. Both the QSH and QAH effects can provide dissipationless conduction channels on the edges like the QH effect, but with no need of external magnetic field, which are useful for low-power electronics, thermoelectrics, topological quantum computation, etc~\cite{qi2010,hasan2010,qi2011,xu2014}.

On the other hand, the fast-growing field of phononics has attracted intensive research interest~\cite{li2012,maldovan2013}, but new ideas and devices are demanded to manipulate phonons efficiently. While in principle phonons cannot have the QSH effect due to the lack of the spin degree of freedom, they are theoretically allowed to have the QH or QAH effect. As learned from electrons, the QH and QAH effects are characterized by a topological invariant called Chern number of the first class, which is defined as an integration of the Berry curvature over the Brillouin zone and can be nonzero only for broken TRS~\cite{thouless1982}. Nontrivial topological states with nonzero Chern number have been well established for photons~\cite{haldane2008,wang2009,lu2014} and recently reported for phonons, acoustic or mechanical systems ~\cite{prodan2009,zhang2010,yang2015,wang2015,kariyado2015,wang2015prl,peano2015,susstrunk2015,nash2015,susstrunk2016,huber2016,liu2017}. As demonstrated by these previous works, topological quantities (like Berry curvature, Berry phase and Chern number) are independent of particle statistics and can be defined for phonons as for electrons. The phononic states with nonzero Chern number give the Q(A)H-like edge states of phonons, which are topologically protected to be gapless, one-way and immune to backscattering~\cite{prodan2009,zhang2010,yang2015,wang2015,kariyado2015,wang2015prl,peano2015,susstrunk2015,nash2015,susstrunk2016,huber2016,liu2017}.

However, it is experimentally more challenging to realize strong TRS breaking effects for phonons than for electrons, since phonons are neutral excitations and couple weakly with magnetic field. Theoretically, a few works proposed to break TRS of phonons by spin-lattice interactions in magnetic materials, Coriolis/magnetic field and optomechanical interactions~\cite{holz1972,strohm2005,sheng2006,zhang2010,wang2015,kariyado2015,peano2015}. But a unified theoretical description of all these interactions is still lacking. Moreover, theoretical approaches are to be developed for investigating quantum transport of phonons with broken TRS. Further study of TRS-breaking and topological effects for controlling phonons in unprecedented new ways is crucial to develop future phononics. Furthermore, fundamental theoretical models like the Kane-Mele model~\cite{kane2005} and the Haldane model~\cite{haldane1988} have been demonstrated to be useful for the study of topological states of electrons~\cite{cayssol2013}. Their phononic analogs are yet to be developed. The exploration of these key issues and related potential applications is of critical importance, since the marriage of topology and phononics could bring in a new paradigm phononics---``topological phononics'', which utilizes novel quantum degrees of freedom like the Berry phase and topological order to control phonons, and may find important applications in heat dissipation, thermoelectrics, thermal insulation, phononic devices, etc~\cite{maldovan2013}.


In this Letter, we propose a Schr{\"o}dinger-like equation of phonons in the extended coordinate-velocity space, so that the influence of TRS is properly included and topological concepts are naturally generalized from electrons to phonons, which offers a general theoretical framework for the study of topological phononics. Then we develop a phononic analog of the Haldane model describing the novel quantum (anomalous) Hall [Q(A)H]-like phonon states. The exotic topological phonon states support one-way gapless edge states immune to scattering. As potential applications, we suggest some unprecedentedly new phononic devices, including low-power phononic circuits, highly efficient phonon valley filters and an ideal phonon diode. Our findings are of great importance to both fundamental research and device applications, shedding new light on the development of topological phononics.

The Lagrangian of a crystal lattice in the harmonic approximation is typically written as
\begin{equation}
L_0 =  \frac{1}{2} \dot{u}_i\dot{u}_i -  \frac{1}{2} D_{ij} u_i u_j  ,
\end{equation}
where $u_i = \sqrt{m_i} x_i$, $m_i$ is the atomic mass, $x_i$ is the displacement, $i,j = 1, 2, ..., dN$ for a $d$-dimensional system including $N$ atoms, $D_{ij} = \frac{\partial^2{V}}{\partial{u_i}\partial{u_j}} |_{{\mathbf{u}}=0}$ is the mass-weighted force constant, and $V$ is the potential energy. Einstein summation convention is used. The equation of motion of lattice vibrations is $\ddot{\mathbf{u}} = - D \mathbf{u}$. Its Fourier transformation gives $D_{{\mathbf{k}}} {\mathbf{u}}_{{\mathbf{k}}}  =  \omega^2_{\mathbf{k}} {\mathbf{u}}_{\mathbf{k}}$. The eigenvalue problem of dynamic matrix $D_{{\mathbf{k}}}$ determines the phonon dispersion $\omega_{\mathbf{k}}$ and wavefunction ${\mathbf{u}}_{\mathbf{k}}$. This is a textbook result, for which TRS is implicitly assumed.

TRS breaking is crucial to realize topological phonon states with nonzero Chern number and to improve efficiency of phonon diode, as shown later. However, its influence on phonons has been rarely studied~\cite{holz1972,strohm2005,sheng2006,zhang2010,peano2015}. Theoretically, the TRS-breaking term that is physically allowed in a harmonic Lagrangian has the form
\begin{equation}\label{L-prime}
L' =  \eta_{ij} \dot{u}_i u_j ,
\end{equation}
where the coefficient matrix $\eta$ is real and antisymmetric, as proved in the Supplementary Material (SM)~\cite{suppl}.

Experimentally, TRS can be broken by (external or internal) magnetic field $\mathbf{B}$. But unlike electrons, neutral phonons do not couple with magnetic field directly. Indirect coupling, however, is possible in ionic lattices by the Lorentz force on charged ions~\cite{holz1972,zhang2010}, or in magnetic lattices by the Raman spin-lattice interaction~\cite{sheng2006,zhang2010} that was used to explain the phonon Hall effect observed experimentally~\cite{strohm2005}.  Moreover, TRS can also be broken by Coriolis field $\mathbf{\Omega}$ introduced through a rotating frame~\cite{wang2015,kariyado2015} or gyroscopic coupling~\cite{wang2015prl}. The Coriolis force $2m \mathbf{v} \times \mathbf{\Omega}$  (mass $m$ and velocity $\mathbf{v}$) has the similar form as the Lorentz force $q \mathbf{v} \times \mathbf{B}$  (charge $q$), thus the two kinds of fields are equivalent. The Lagrangian of $\mathbf{\Omega}$ is described by Eq.~\eqref{L-prime}, for which the submatrix $\hat \eta_{\mu \nu}$ of $\eta$ between atoms $\mu$ and $\nu$ is
\begin{equation}
\hat \eta_{\mu \nu} =  \delta_{\mu \nu} \left(
  \begin{array}{ccc}
    0 & -\Omega_z & \Omega_y \\
    \Omega_z & 0 & -\Omega_x \\
    -\Omega_y & \Omega_x & 0 \\
  \end{array}
\right),
\end{equation}
where $\delta_{\mu \nu}$ is the Kronecker delta function. The Lagrangian of $\mathbf{B}$ is obtained correspondingly by replacing $\mathbf{\Omega}$ with $q {\mathbf B}/2m$.

When the TRS-breaking term $L'$ is included, the equation of motion of lattice vibrations becomes $\ddot{\mathbf{u}} = - D \mathbf{u} - 2\eta \dot{\mathbf{u}}$. Its Fourier transformation changes into $D_{{\mathbf{k}}} {\mathbf{u}}_{{\mathbf{k}}} - 2i \omega_{{\mathbf{k}}} \eta_{\mathbf{k}}  {\mathbf{u}}_{\mathbf{k}} =  \omega^2_{\mathbf{k}} {\mathbf{u}}_{\mathbf{k}}$. Noticeably, this equation is no longer an eigenvalue problem (since $\eta_{\mathbf{k}}$ is a matrix), and thus the solution ${\mathbf{u}}_{\mathbf{k}}$ can not be viewed as eigenmode anymore. Some fundamental problems arise: How to properly define the wavefunction, Berry phase and topological invariants for phonons? How to describe the effects of TRS breaking in quantum phonon transport?

Let us borrow ideas from electrons, following the work of Ref.~\onlinecite{kane2014}. If we can find a Schr{\"o}dinger-like equation for phonons, all the above problems would be naturally solved. However, this is not a straightforward task, since the Schr{\"o}dinger equation of electrons is first order in time derivative and has positive and negative energies, whereas the Lagrangian equation of phonons is second order in time and has non-negative frequencies only~\cite{kane2014}. Actually, a similar problem has been solved by Dirac, who discovered the Dirac equation by taking a ``square root'' of the Klein-Gordan equation. As inspired by Dirac, we take a ``square root'' of the Lagrangian equation by reformulating the problem of lattice vibrations with Hamiltonian mechanics, which gives a \textit{Schr{\"o}dinger-like equation of phonons}: $H_{\mathbf k} {\mathbf \psi_{\mathbf k}} = \omega_{\mathbf k} {\mathbf \psi_{\mathbf k}}$, where
\begin{equation}\label{Hamiltonian}
H_{\mathbf k} = \left(
\begin{array}{ccc}
0                      & i D_{\mathbf k}^{1/2} \\
-i D_{\mathbf k}^{1/2}     & -2i{\eta}_{\mathbf k}
\end{array}
\right),~~~
{\mathbf \psi_{\mathbf k}} = \left(
\begin{array}{c}
  D_{\mathbf k}^{1/2} {\mathbf u_{\mathbf k}} \\
  {\dot{\mathbf{u}}_{\mathbf k}}
\end{array}\right).
\end{equation}
Note that $D_{\mathbf k}$ is semi-positive-definite as required by the structural stability of the system, and $D_{\mathbf k}^{1/2}$ can be constructed by the eigenvalues and eigenvectors of $D_{\mathbf k}$. Thus the Hamiltonian can be built with eigenvalues and eigenvectors of $D_{\mathbf k}$. Based on the extended coordinate-velocity space, the phonon wavefucntion and all topology-related quantities (like Berry curvature ${\mathbf B}_{\mathbf k}$ and Chern number  $\mathcal{C}$) can be defined similar as for electrons~\cite{suppl}, which was discussed previously by using a non-Hermitian Hamiltonian~\cite{zhang2010,suppl}. The equation has an intrinsic ``particle-hole'' symmetry that guarantees $(\omega_{\mathbf k}, {\mathbf \psi}_{\mathbf k})$ and $(-\omega_{-\mathbf k},{\mathbf \psi}_{\mathbf k}^*)$ appear in pairs. Under inversion symmetry, $\omega_{\mathbf k} = \omega_{-\mathbf k}$ and ${\mathbf B}_{\mathbf k}={\mathbf B}_{-\mathbf k}$. Under TRS, $\omega_{\mathbf k} = \omega_{-\mathbf k}$, ${\mathbf B}_{\mathbf k}=-{\mathbf B}_{-\mathbf k}$ and $\mathcal{C} \equiv 0$. In the following, we first present a general discussion on the differences between the problems of phonons and electrons, and then generalize the nonequilibrium Green's function (NEGF) method~\cite{xu2008,wang2008,xu2010} to study quantum phonon transport with broken TRS.

Generally, the problem of phonons differs significantly from that of electrons at least on the following four aspects:

(i) \textbf{Hamiltonian and basis orbital}: The (tight-binding) Hamiltonians of electrons and phonons are built with basis orbitals of different orbital functions and dimensions. For electrons, basis orbitals are selected according to the chemical nature of the composed elements, and the number of basis orbitals ($n$ per atom) is flexible for varying purpose of simplicity and accuracy. For instance, the $p_z$ orbital is usually selected as the basis for graphene, and a more sophisticated study would include the $s$ and $p_{x,y,z}$ orbitals into the basis. The site-site coupling is then described by a $n\times n$ matrix and the Hamiltonian $H_{\mathbf k}$ by a $nN\times nN$ matrix ($N$ atoms per unit cell). For phonons in a $d$-dimensional space with (without) time reversal symmetry (TRS), the interatomic coupling is described by a $d\times d$ ($2d\times 2d$) matrix and $H_{\mathbf k}$ has a fixed dimension of $dN \times dN$ ($2dN \times 2dN$). The underlying basis orbitals, from the point view of symmetry, are $p_{x,y,z}$ for $d=3$, $p_{x,y}$ for $d=2$ and $p_x$ for $d=1$. Moreover, $H_{\mathbf k}$ of phonons satisfies the acoustic sum rule due to the rigid translational symmetry, implying that its on-site elements cannot be tuned freely as for electrons.

(ii)\textbf{ Wavefunction}: Wavefunctions of phonons are vectors in the coordinate-velocity space, showing symmetries properties different from those of electrons. Specifically, the Bloch wavefunctions of phonons under space inversion and time reversal operations will not only change the indices (i.e., the $K$/$K'$ valley, wavevector $\mathbf k$ and sometimes $A$/$B$ sublattice) like for electrons, but also introduce additional changes: the coordinate part changes sign under space inversion; the velocity part changes sign under space inversion and time reversal.

(iii) \textbf{Particle statistics}: Unlike electrons, phonons as bosons are not limited by the Pauli exclusion principle, implying that the whole phonon spectrum can be physically probed. In this work we focused on discussing topological properties of the band gap close to the Dirac point. Other band gaps, like the one between the two uppermost bands, could also be topologically nontrivial [see Fig. 1(d)]. All these nontrivial band gaps support the Q(A)H-like phonon states. Topological effects of phonons can be observed for a given frequency and a given wavevector, for instance, by Raman or infrared experiments. The nontrivial topological states can be detected by measuring phonons of frequencies within the bulk band gap, which are absent in the bulk but appear on the edges/boundaries.

(iv) \textbf{Magnetic translation}: For electrons under a magnetic field, translation operators along different directions usually do not commute due to the existence of the Aharonov-Bohm phase. Thus magnetic Bloch states and magnetic Brillouin zone are used to describe magnetic fields~\cite{xiao2010}. In contrast, we are able to use the usual Bloch states and Brillouin zone to study phonons under magnetic (or TRS breaking) fields, because phonons as neutral elementary excitations do not have the Aharonov-Bohm phase. Importantly, TRS-breaking fields break the rigid translational symmetry and the acoustic sum rule, which splits the acoustic bands at the $\Gamma$ point [see Fig. 1(d)].

For phonon transport, the NEGF method has been widely used to study the problem quantum mechanically, which is able to describe linear phonon systems accurately and provide a systematic way to treat many-body interactions~\cite{xu2008,wang2008,xu2010}. However, the existing NEGF method works only for phonons with TRS. The approach has to be generalized to describe phonons with broken TRS. Here based on the Schr{\"o}dinger-like equation of phonons, we derive a generalized formalism of the NEGF method for exploring the effects of TRS breaking in quantum phonon transport.

An infinite transport system has infinite number degrees of freedom. To make the problem solvable, the transport system in theoretical calculations is usually divided into three parts: the center (C) part where phonon scattering happen, together with semi-infinite leads on the left (L) and right (R), which act as reservoir of phonons. The influence of reservoirs are projected into the center part through self energies $\Sigma_{L}$ and $\Sigma_{R}$.

One may define a Green's function $G^{r(a)}(\omega)=[(\omega \pm i0^+)  - \mathcal{H}]^{-1}$ in analogy to electrons, where $G^{r}(\omega)$ and $G^{a}(\omega)$ are retarded and advanced Green's functions, respectively. However, interatomic couplings in this picture are not short-range due to the existence of the $D^{1/2}$ term in $\mathcal{H}$. Therefore, we cannot decouple the left and right semi-infinite leads, making practical calculations difficult. The problem is solved by changing the phonon equation into a generalized Hermitian eigenvalue problem $Q \tilde{y} = \omega R^{-1} \tilde{y}$, where $R$ and $Q$ are Hermitian matrices defined by $D$ and $\eta$ (see details in the SM~\cite{suppl}). Thus we can avoid using $D^{1/2}$ and define a new Green's function:
\begin{equation}
G^{r(a)}(\omega)=[(\omega \pm i0^+) R^{-1} - Q]^{-1}.
\end{equation}
In this new picture, interatomic couplings as described by $D$ are typically short-range. Thus we can calculate self energies $\Sigma_{L}$ and $\Sigma_{R}$ of the semi-infinite leads iteratively as done previously for systems with TRS~\cite{wang2008,xu2010}. The key quantity to be calculated is the Green's function of the center part $G_C$. For simplicity the subscript ``C'' is omitted hereafter. $G^{r(a)}$ of the center part can be calculated  by the Dyson equation:
\begin{equation}
G = G_0 + G_0 (\Sigma_L + \Sigma_R) G,
\end{equation}
where $G_0$ is the free Green's function of the center part decoupled from the reservoirs. With the calculated Green's function, any single-particle quantities can be obtained~\cite{wang2008,xu2010}, including the phonon transmission function from the left lead to the right lead, which is given by
\begin{equation}
\Xi(\omega) = \mathrm{Tr}(G^r\Gamma_RG^a\Gamma_L),
\end{equation}
where $\Gamma_\alpha=i(\Sigma^r_\alpha-\Sigma^a_\alpha)$~($\alpha=L, R$).

\begin{figure}
\centering
\includegraphics[width=\linewidth]{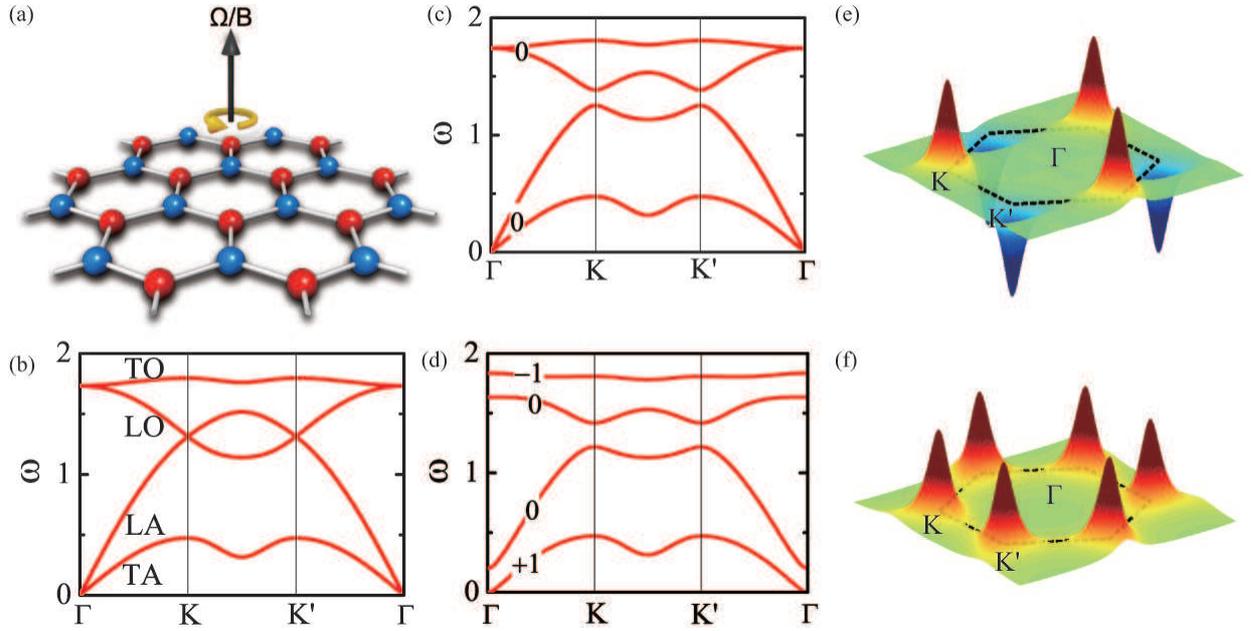}
\caption{\label{fig1}
(a) A two-dimensional honeycomb lattice whose inversion symmetry is broken by different atomic masses of the two sublattices (colored red and blue) and TRS is broken by Coriolis field ${\mathbf \Omega}$ or magnetic field ${\mathbf B}$. (b) The phonon dispersion for no symmetry breaking ($m_I=m_T=0$), where Dirac-like linear dispersions occur at the $K$ and $K'$ points. ``TA'', ``LA'',``TO'' and ``LO'' denote transverse acoustic, longitude acoustic, transverse optical, and longitude optical modes, respectively. (c,d) The Phonon dispersions for (c) broken inversion symmetry ($m_I=0.1$) and (d) broken TRS ($m_T=0.1$), where a band gap is opened near the Dirac point. Chern numbers of separated bands are labelled. The same values of Chern numbers were obtained by a different computational method in a previous work~\cite{wang2015}. (e,f) Distribution of the Berry curvature in the Brillouine zone for the two lowest bands displayed in (c) and (d), respectively.}
\end{figure}

Next we apply the developed theory to study the in-plane vibrations of a two-dimensional honeycomb lattice [Fig.\ref{fig1}(a)], a model system used for investigating the QSH and QAH effects of electrons~\cite{haldane1988,kane2005}. The in-plane vibrations of the honeycomb lattice with broken TRS and the appearance of QAH-like states of phonons have been studied previously by using different theoretical methods~\cite{zhang2010,wang2015,kariyado2015}. The influence of out-of-plane vibrations is discussed in the SM~\cite{suppl}. Interatomic couplings with nearest- and next-nearest neighbors are included and described by springs of force constants $k_1$ and $k_2$, respectively. Inversion symmetry is broken by using different atomic masses of the two sublattices $m_A = m(1 - \delta)$ and $m_B = m(1 + \delta)$ as tuned by $\delta$. TRS is broken by Coriolis field induced through a rotating frame with out-of-plane angular velocity $\Omega_z$ or equivalently by magnetic field. $k_1 = 1$, $k_2 = 0.05$, $m=1$ and the nearest neighbor distance $a = 1$ were used without loss of generality. Typical phonon dispersions are presented in Figs.\ref{fig1}(b-d). For no symmetry breaking, the longitudinal acoustic (LA) and longitudinal optical (LO) bands form Dirac-like linear dispersions at the $K$ and $K'$ points [Fig.\ref{fig1}(b)]. The Dirac points, located at $\omega_D = \sqrt {(3 k_1 + 9 k_2)/2m }$ with group velocity $v_D=3k_1a/8m\omega_D$, are protected by inversion symmetry and TRS. A band gap opens at the Dirac point when either symmetry is broken [Figs.\ref{fig1}(c,d)]. We name it the Dirac gap and discuss its topological nature below.

To study the interplay of symmetry and topology, we derive an effective Hamiltonian for phonons near the Dirac point based on symmetry analysis~\cite{suppl}:
\begin{equation}\label{Effective_H}
H_{0}({\mathbf k}) = v_D\left(k_y\tau_z\sigma_x - k_x\sigma_y\right) ,
\end{equation}
where ${\mathbf k}$ is referenced to $K$ ($K'$) and $\omega_{\mathbf k}$ to $\omega_D$. $\sigma$ and $\tau$ are the Pauli matrices with $\sigma_z=\pm1$ and $\tau_z=\pm1$ referring to $A(B)$ sublattice and valley index $K$ ($K'$), respectively. This Hamiltonian gives the linear gapless phonon dispersions $\omega_{\mathbf k} =  \pm v_D|{\mathbf k}|$ near the $K$ and $K'$ points. The same Hamiltonian has been used to describe the Dirac electrons of graphene~\cite{hasan2010,cayssol2013}.

When breaking inversion symmetry, a mass term is added into the effective Hamiltonian
\begin{equation}\label{HI}
H_I^\prime = m_I\sigma_z,
\end{equation}
where $m_I={\displaystyle \frac{\delta}{2}}\omega_D$~\cite{suppl}. This is a Semenoff model~\cite{semenoff1984} for phonons. The Semenoff-type mass term has minor effect on the whole phonon dispersion except around the Dirac point, where $\omega_{\mathbf k} =  \pm \sqrt{ (v_D|{\mathbf k}|)^2 + m_I^2 }$ and a band gap of $2 m_I$ is opened at the Dirac point [Fig.\ref{fig1}(c)].

When breaking TRS, another kind of mass term is added into the effective Hamiltonian
\begin{equation}\label{Haldane}
H_T' = m_T\sigma_z\tau_z,
\end{equation}
where $m_T = \Omega_z$~\cite{suppl}. This gives $\omega_{\mathbf k} =  \pm \sqrt{ (v_D|{\mathbf k}|)^2 + (m_T \tau_z)^2 }$, opening a band gap of $2m_T \tau_z$ at the Dirac point [Fig.\ref{fig1}(d)]. This band gap has opposite signs at the $K$ and $K'$ points, indicating an inverted band order. To smoothly transform into trivial insulators in the atomic limit, a band inversion must happen and the band gap must be closed at the critical transition point. In contrast, the band gap generated by $H_I'$ has the same sign at the $K$ and $K'$ points, which can be adiabatically connected to trivial atomic insulators. Therefore, the Dirac gap induced by $H_T'$ and $H_I'$ belong to two different topological classes. One is topologically nontrivial, the other is trivial.

Importantly, Eqs.~\eqref{Effective_H} and ~\eqref{Haldane} are reminiscent of the well-known Haldane model~\cite{haldane1988,kane2005}, the first model of the QAH effect of electrons. This exotic quantum effect has recently been experimentally observed in a magnetic-doped TI~\cite{chang2013}. We propose a phononic analog of the Haldane model, which gives the novel Q(A)H-like phonon states, as we will demonstrate. More advantageous than the original Haldane model~\cite{haldane1988,elder2013}, this model requires no complicated magnetic flux and does not give a vanishingly small band gap caused by local orbital constraints~\cite{suppl}.

\begin{figure}
\centering
\includegraphics[width=\linewidth]{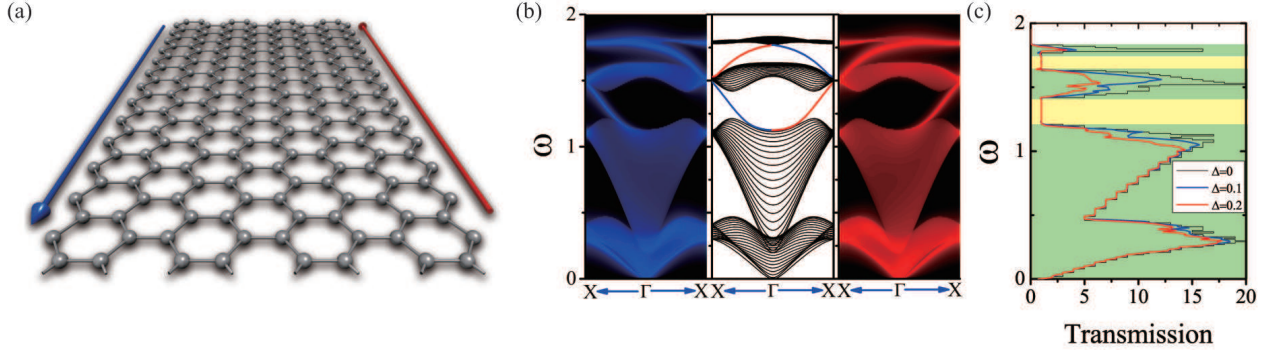}
\caption{\label{fig2}
(a) Schematic of a nanoribbon structure with one-way edge states. A nanoribbon including 20 zigzag chains is calculated with broken TRS ($m_T = 0.1$). (b) The phonon dispersion (middle panel), and density of states projected onto the left/right edges (left/right panels), where large (small) values are colored blue/red (black). The backward-moving (blue) and forward-moving (red) edge modes appear only on the left and right, respectively. (c) Phonon transmissions as a function of frequency $\omega$ for varying strength of disorder $\Delta$. The disorders are introduced into the center part of a length of 20 unit cells by selecting atomic masses randomly within $[1 - \Delta, 1 + \Delta]$. Results of $\Delta = 0, 0.1, 0.2$ are denoted by black, red and blue lines, respectively. The phonon transmission decreases gradually with increasing $\Delta$, except for the one-way edge states within the bulk gap (shaded yellow).}
\end{figure}

To verify the topological nature, we calculate the Berry curvature and Chern number of the Dirac gap, as shown in Fig.~\ref{fig1}. Moreover, we compute the phonon band structure for a nanoribbon structure with $m_T = 0.1$, showing the existence of the one-way edge phonon modes [Figs.~\ref{fig2}(a,b)], as investigated before in previous works~\cite{wang2015,kariyado2015}. Furthermore, we perform transport calculations by the generalize NEGF method~\cite{suppl} to explore disorder effects, which clearly proves that the one-way edge states remain ballistic irrespective of the disorder [Fig.~\ref{fig2}(c)]. More details are described in the SM~\cite{suppl}.

\begin{figure}
\centering
\includegraphics[width=\linewidth]{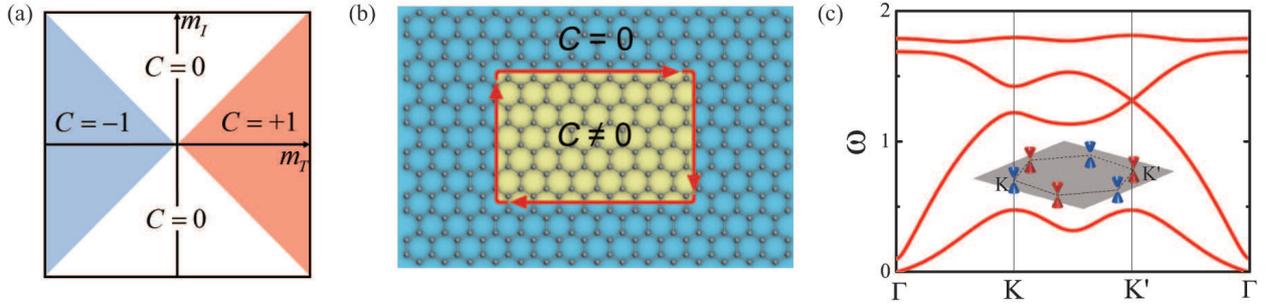}
\caption{\label{fig3}
(a) A phase diagram showing that the Chern number $\mathcal{C}$ of the Dirac gap is tuned by the two mass terms $m_I$ and $m_T$. (b) Schematic of patterning one-way edge states (denoted by red lines) at the interface between regions of $\mathcal{C} = 0$ and $\mathcal{C} \ne 0$. (c) The phonon dispersion of $m_I=m_T=0.05$, in which one Dirac point at the $K'$ valley is preserved and the other one at the $K$ valley is gapped as illustrated in the inset. This is an example showing that the band degeneracy of the $K$ and $K'$ valleys is split by breaking inversion symmetry and TRS simultaneously.}
\end{figure}

Remarkably, the valley degree of freedom can be significantly tuned by breaking inversion symmetry and TRS simultaneously, offering new opportunities for valley phononics. Specifically, the two Dirac gaps at the $K$ and $K'$ valleys change independently when varying $m_I$ and $m_T$: One is $2(m_I+m_T)$, the other is $2(m_I-m_T)$. As an extreme case, a novel phonon dispersion having a Dirac gap at the $K$ valley and a gapless Dirac cone at the $K'$ valley is obtained by selecting $m_I = m_T \ne 0$ [Fig.~\ref{fig3}(c)]. A highly efficient valley-polarized phonon current could be realized by passing phonons of frequencies within the Dirac gap of the $K$ valley through this structure, which acts as an ideal phonon valley filter. This finding could find important applications in future phononic devices, noticing that valley phonons can be selectively excited by polarized photons and detected by a valley phonon Hall effect~\cite{zhang2015}.

Furthermore, the Dirac gaps have the same sign between $K$ and $K^\prime$ when $|m_T| < |m_I|$, giving $\mathcal {C} = 0$; their signs are opposite when $|m_T| > |m_I|$, giving $\mathcal {C} = 1$. The competition between $m_I$ and $m_T$ determines the band topology of the Dirac gap, as summarized in the phase diagram of Fig.~\ref{fig3}(a), which provides a guidance for designing topological phononics. As a potential application, the one-way edge states, which can act as dissipationless conducting wires of phononic circuits, could be patterned in a controlled way as illustrated in Fig.~\ref{fig3}(b). The lattice boundaries are designed with varying $m_I$ ($m_{I1}$ and $m_{I2}$) and a homogenous TRS-breaking field $m_T$. The one-way edge states exist at the boundary when $|m_{I1}| < |m_T| < |m_{I2}|$ and disappear otherwise. We performed molecular dynamics simulations to visualize the lattice dynamics, and confirmed that lattice vibrations of frequencies within the Dirac gap transport unidirectionally along the patterned boundary. The calculation details and snapshots of a movie are included in the SM~\cite{suppl}.

\begin{figure}
\centering
\includegraphics[width=\linewidth]{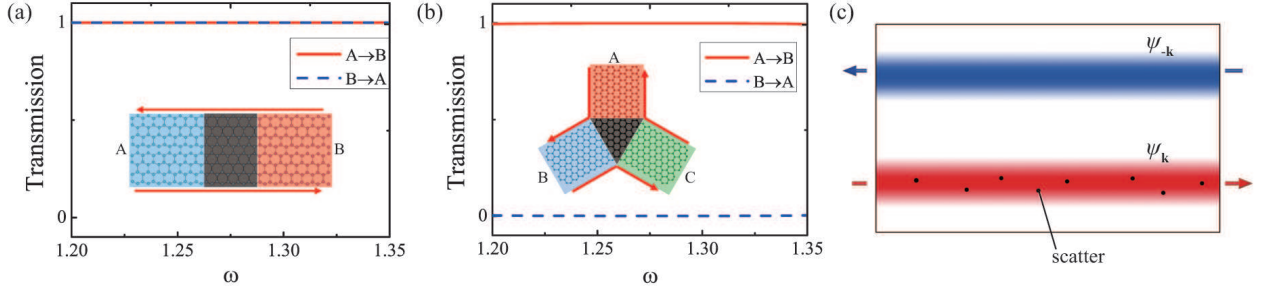}
\caption{\label{fig4}
(a,b) Phonon transmission functions for transport from the terminal A to B (red solid line) and from B to A (blue dashed line) in two-terminal and three-terminal transport systems ($m_I=0, m_T=0.1$). Results are shown only for phonon frequencies within the Dirac gap, where the one-way edge states (denoted by red lines in the insets) exist. The former system gives zero diode effect, while the latter acts as an ideal phonon diode. (c) A schematic diagram illustrating the concept of selective scattering for enhancing the diode effect. The real-space distribution of the forward-/backward-moving states (${\mathbf \psi}_{\mathbf k}$/${\mathbf \psi}_{- \mathbf k}$) (denoted by red/blue colors) could be separated in real space when breaking TRS. Then, scatters (denoted by black dots) can be selectively placed on the transport path of the forward-moving state, introducing significant scattering for ${\mathbf \psi}_{\mathbf k}$ but not ${\mathbf \psi}_{- \mathbf k}$.}
\end{figure}

Another promising application of the one-way edge states is phonon diode, a key component of phononics, which conducts phonons in one direction but blocks phonons in the opposite direction. A general theorem tells that the diode effect is zero in linear lossless systems with TRS for waves of any nature~\cite{maznev2013}. Thus one may design phonon diode based on nonlinearity or TRS breaking. Previous research mainly tried the way of nonlinearity, obtaining extremely low efficiencies~\cite{li2004,chang2006}. This could be explained by the fact that the forward- and backward-moving states would always appear in pairs when having TRS. We suggest to break TRS for disentangling the forward- and backward-moving states, thus improving the diode effect. The unique one-way edge states perfectly fit the purpose.

To see diode effect explicitly, we considered a typical two-terminal transport system and simulated phonon transport by the generalized NEGF method~\cite{suppl} [Fig.~\ref{fig4}(a)]. Unfortunately, the calculated diode effect is always zero for both the bulk and edge states, despite trying different parameters and disorders. Actually, this conclusion generally applies to two-terminal coherent transport, explained by the unitary condition of the scattering matrix (proved in the SM~\cite{suppl}). This is straightforward for the one-ways edge states, whose diode effect gets cancelled by the two opposite edges. Then we turn to study a three-terminal transport system, which was investigated previously for an acoustic isolator~\cite{fleury2014}. There, the one-way edge states is able to give perfect diode effect between any two of the terminals. For instance, the edge states flow from the terminal $A$ to $B$ without any scattering, and the back flow from $B$ to $A$ is exactly zero (the input from $B$ all flows out through $C$). This physical picture is verified by the transport calculations [Fig.~\ref{fig4}(b)]. An ideal phonon diode is thus realized by the one-way edge states in a multi-terminal setup. As far as we know, this is the first model of ideal phonon diode, which is promising for future phononics.

For generic phonon states, we propose a concept of selective scattering to enhance the diode effect based on TRS breaking [illustrated in Fig.~\ref{fig4}(c)], which are described in the SM~\cite{suppl} and will be explored in our future work. Our findings of topological phononics and related device applications could also be applied to control sound and heat, although the manipulation of heat conduction is relatively more challenging because all the thermally excited phonons are involved in the process.

Many two-dimensional materials are known to be Dirac electronic materials, including graphene, silicene, germanene, stanene, graphynes, several boron and carbon sheets, etc.~\cite{wang2015nsr}. These materials are very likely to have linear Dirac-like phonon dispersions. Breaking TRS of their phonons, for instance, by spin-lattice interactions in magnetic materials, Coriolis/magnetic field or light-matter interactions, could open a nontrivial phonon band gap at the Dirac point, whose magnitude $\Delta$ is proportional to the strength of TRS breaking. For example, $\Delta = 2 \Omega_z$ for the Coriolis field generated by a rotation angular velocity of $\Omega_z$. An important subject for future research is to find efficient ways/mechanisms to realize strong TRS breaking effects of phonons in solids.

\emph{Note added}: Another group derived a different version of Schr\"{o}dinger-like equation for phonons while we were finalizing this manuscript~\cite{susstrunk2016}.

\begin{acknowledgements}
Y.X. acknowledges support from the National Thousand-Young-Talents Program and Tsinghua University Initiative Scientific Research Program. Y.L. and W.D. acknowledge support from the Ministry of Science and Technology of China (Grant No. 2016YFA0301001) and the National Natural Science Foundation of China (grant nos. 11674188 and 11334006). S.C.Z is supported by the NSF under grant numbers DMR-1305677.
\end{acknowledgements}


\end{document}